\documentstyle[11pt,newpasp,twoside]{article}
\begin{document}

\title{Polarization Characteristics of Pulsar Profiles}

\author{J.L. Han$^{1,2}$, R.N. Manchester$^2$, G.J. Qiao$^3$}

\affil{$^1$National Astronomical Observatories, CAS, Beijing 100012, China\\
       $^2$Australia Telescope National Facility, CSIRO, Australia\\
       $^3$Department of Astronomy, Peking University, Beijing 100871, China}

\begin{abstract}
Polarization profiles of several hundred pulsars have been published
recently (eg. Gould \& Lyne 1998, GL98; Weisberg et al. 1999).
In this report, we
summarize the characteristics of circular and high linear polarizations
of pulsar profiles, based on all previously published data.
\end{abstract}

\vspace{-8mm}
\section{Circular Polarization}
Han et al. (1998) systematically studied the circular polarization of
pulsar profiles. They found that circular polarization is common in
profiles but diverse in nature. Circular polarization is about 9\% in
average, weaker than linear polarization (15\%) in general. We emphasize
the following points about circular polarization (CP):

\noindent {\bf (1). CP not unique to core emission:} 
One misleading concept is that circular polarization alway accompanies
core emission. It is generally strongest in the central or
`core' regions of a profile, but is by no means confined to these regions.
Circular polarization has been detected from conal components of many
pulsars, for example, conal-double  pulsars.

\noindent {\bf (2). Sense reversal not unique to core:}
Circular polarization often changes sense near the middle of the profile.
But sense reversals have been observed at other longitudes,
e.g. conal components of PSRs B0834+06, B1913+16, B2020+28,
B1039$-$19, J1751$-$4657, and B0329+54 in abnormal mode. 

\noindent {\bf (3). No PA correlation for core emission:}
There is no correlation between the sense of the sign change of circular
polarization and the sense of variation of linear polarization angle (PA), in
contrast to earlier conclusions on this issue.

\noindent {\bf (4). Correlation for cone-dominated pulsars:}
We found a strong correlation between between the sign of PA variation
and sense of circular polarization in conal-double pulsars, with right-hand
(negative) circular polarization accompanying increasing PA and vice versa.
No good examples contrary to this trend have been found.

\noindent {\bf (5). Variation with frequency:} Circular polarization
generally does not vary systematically with frequency. Multifrequency
profiles of many pulsars show similar CP across a wide frequency range
(eg. PSRs B0329+54, B0525+21). However, we have found several examples with
significant variations (Table 5 of Han et al 1998). For example, PSR
B1749$-$28 now has been confirmed by the data of GL98 to change from dominant
right hand (``$-$'') CP at low frequencies to a sense change ``$-/+$'' at
high frequencies, while PSRs B1859+03 and B1900+01 change from ``$+$'' to
``$+/-$''. 

\section{High Linear Polarization}

Strong linear polarization is an outstanding characteristics of pulsars. 
Almost all pulsars with log $\dot{E} > 34.5$, if a polarization profile
is available, have (at least) one highly linearly polarized component.
Using our profile database, we found that the highly linearly polarized 
components do not have to be associated with high $\dot{E}$. 
Several types of pulsar profiles have highly polarized components:

\noindent {\bf (1). Leading-polarized component:}
The prototype of this kind of pulsars is PSR B0355+54. The leading component
is almost 100\% polarized. The component does not dominate the profile below
several hundred MHz, but becomes stronger towards high frequency. 
%According to the polarization angle curve, very probably this component
%is not emitted from the same region as other part of profiles.
It may be emitted from a different region. Other examples
are PSRs B0450+55, B1842+14, B2021+51, B0809+74, B0626+24, B1822$-$09.

\noindent {\bf (2). Trailing-polarized component:}
PSR B0559$-$05 is a mirror-symmetrical type to PSR B0355+54. Its trailing
component is highly linearly polarized, and becomes stronger with increasing
frequency. Good examples are PSR B2224+65, J1012+5307 and J1022+1001.
The latter two are millisecond pulsars.

\noindent {\bf (3). Polarized multicomponents:}
The prototype PSR B0740$-$28 has 7 Gaussian components fitted to
a high time resolution profile. These pulsars have a sharp
leading edge and more gradual trailing edge, and are highly
polarized. Good examples are PSRs J0538+2817, B0540+23, B0833$-$45,
 B0950+08i, J1359$-$6038, B1929+10m. 
PSRs B0538$-$75, J0134$-$2937 and the postcursor of B0823+26
have just mirror-symmetrical profiles to these examples above.

\noindent {\bf (4). Polarized single component:} Almost 100\% polarized
single components may be conal emission emitted
from the very outer edge of beam. Two dozen examples have been found.
The best examples are PSRs B0105+65, B0611+22, B0628$-$28, B1322+83,
B1556$-$44, J1603$-$5657, B1706$-$44, B1828 $-$10, B1848+13, B1913+10,
B1915+13.

\noindent {\bf (5). Interpulse pulsars:} PSR B0906$-$49 and B1259$-$63
are young, energentic ($\dot{E}>35$), interpulse pulsars with extremely
highly polarization. These may be pulsars where both sides of  a wide
conal beam from a single pole are observed. 
%Further evidence comes from another young pulsar 
PSR J0631+1036 may be an off-centre cut through a wide cone.
%PSRs B1800$-$21, B1823$-$13 have just the first half of the profile
%of J0631+1036.

\medskip

The polarization characteristics of the mean pulse profile provide a
framework for understanding the emission processes in pulsars. 
The characteristics of pulsar circular polarization summarized by Han et
al. (1998) should be considered by all emission models. The points on highly
linear polarization we make here should be included into future pulsar
classifications and geometrical studies of pulsar emission beam.

\medskip

{\bf Acknowledgements \ } JLH thanks financial support from the National
Natural Science Foundation (NNSF) and the Educational Ministry of China.


\begin{references}
\reference Gould D.M., Lyne A.G., 1998, MNRAS 301, 235
\reference Han J.L., Manchester R.N., Xu R.X., Qiao G.J., 1998, MNRAS 300, 373
%\reference Manchester R.N., Han J.L., Qiao G.J., 1998, MNRAS 295, 280
\reference Weisberg J.M., Cordes J.M., Lundgren S.C., et al., 1999, ApJS
121, 171
\end{references}
\end{document}